\documentclass [12pt,preprint]{aastex}
\usepackage{epsfig}
\begin{document}
\voffset-1cm
\newcommand{\gsim}{\hbox{\rlap{$^>$}$_\sim$}}
\newcommand{\lsim}{\hbox{\rlap{$^<$}$_\sim$}}

\title{The time ending the shallow decay \\
of the X-ray light curves of long GRBs}

\author{Shlomo Dado\altaffilmark{1}, Arnon Dar\altaffilmark{1}
and A. De 
R\'ujula\altaffilmark{2}}

\altaffiltext{1}{dado@phep3.technion.ac.il, arnon@physics.technion.ac.il,
dar@cern.ch.\\
Physics Department and Space Research Institute, Technion, Haifa 32000,
Israel}
\altaffiltext{2}{alvaro.derujula@cern.ch; Theory Unit, CERN,
1211 Geneva 23, Switzerland \\ 
Physics Department, Boston University, USA}

\begin{abstract} 
We show that the mean values and distributions of the 
time ending the shallow decay of the light curve of the X-ray afterglow of 
long gamma ray bursts (GRBs), the equivalent isotropic energy in the X-ray 
afterglow up to that time and the equivalent isotropic GRB energy, as well 
as the correlations between them, are precisely those predicted by the 
cannonball (CB) model of GRBs. Correlations between prompt and 
afterglow observables are important in that they test the overall consistency
of a GRB model. In the CB model, the prompt and afterglow 
spectra, the endtime, the complex canonical shape of the X-ray afterglows 
and the correlations between GRB observables are not surprises, but 
predictions.

\end{abstract}

\section{Introduction}
\label{Introduction}

In a relatively brief time, the observations made or triggered by the 
Burst Alert Telescope (BAT), the X-Ray Telescope (XRT), and the UVOR 
telescope, aboard the Swift satellite, have gathered a wealth of new 
information on Gamma Ray Bursts (GRBs), in particular on the early X-ray 
and optical afterglows (AGs) of long-duration GRBs. These observations 
pose severe problems to the generally accepted `Fireball' model of GRBs 
(see, e.g.~Meszaros 2006), whose `microphysics' (see, e.g.~Panaitescu et 
al.~2006), reliance on shocks (see, e.g.~Kumar et al.~2007), and 
correlations based on the `jet-opening angle' (see, e.g.~Sato et al.~2007;
Burrows \& Racusin~2007),  may have to be abandoned.

The said recent observations agree remarkably well with the 
predictions of the `Cannon Ball' (CB) model (Dar \& De R\'ujula~2004;
Dado, Dar \& De R\'ujula~2002; Dado, Dar \& De R\'ujula~2003, hereafter
DD04; DDD02; DDD03, respectively). Some examples are given in Fig.~\ref{f1}. 
The predicted lightcurve of the X-ray AG afterglow is shown in 
Fig.~\ref{f1}a for the fireball (see e.g.~Maiorano et al.~2005) and CB 
(DDD02) models. For GRBs with 
an approximately constant circumburst density distribution, the 
well-observed X-ray AGs have a `canonical behaviour' (e.g.~Nousek et 
al.~2006; Zhang et al.~2006; O'brein et al.~2006), in impressive 
agreement with the CB-model predictions, as in the example of 
Fig.~\ref{f1}b. The evolution of the AG around the time, $T_a$, ending the 
`shallow phase' is predicted to be achromatic in the optical to X-ray 
range (DDD02), as can be seen in Fig.~\ref{f1}c and its comparison with 
Fig.~\ref{f1}d. This `achromaticity' does not extend to the radio domain 
(DDD03). Many correlations 
between (prompt) GRB observables have been studied with the help of the 
new data on GRBs of measured red-shift $z$ (e.g. Schaefer~2006). All of 
the successful correlations are simple predictions of the CB model. One 
example, perhaps the best known, is the correlation between the spectral 
`peak' energy and the total (bolometric) isotropic-equivalent energy, 
shown in Fig.~\ref{f1}d along with our prediction (Dado at al.~2007 and 
references therein).

Willinger et al.~(2007) have studied and tabulated recent data on $T_a$, 
the time ending the shallow decay of the X-ray light curves of long GRBs. 
In a paper the title of whose first version was the same as ours, 
Nava et al.~(2007) have investigated the correlation between $T'_a\equiv 
T_a/(1+z)$, the endtime in the source's rest frame, and the prompt 
GRB energy, as well as the correlation between $T'_a$ and the energy in the 
X-ray plateau phase, integrated up to $T_a$. The values and distributions 
of these three quantities, as well as their correlations, are important in 
ascertaining the global validity of a GRB model, among other things 
because they test the consistency of the description of the prompt and 
afterglow phases. In this paper we show that the observations are in 
precise agreement with the CB-model's predictions. To do so, we gather, in 
Sections \ref{dectime}, \ref{promptE} and \ref{XAGenergy}, predictions 
from various of our papers, in a manner and order adequate to the 
discussion of the subject at hand. The predictions are derived for 
`typical' or average values of the parameters, all chosen as in our 
earlier work, which referred mainly to pre-Swift observations. The 
incidence and explicit origin of the variability around the typical cases 
is also discussed. The results are summarized in Fig.~\ref{f4}, which 
demonstrates that the central expectation, variability and correlations 
are all as predicted.

\section{The CB model}
\label{CBmodel} 

In the CB model (Dar \& De R\'ujula~2000, DD2004; DDD02, 
DDD03), {\it long-duration} GRBs and their AGs are produced by bipolar 
jets of CBs, ejected in core-collapse SN explosions (Dar \& Plaga~1999). 
An accretion disk is hypothesized to be produced around the newly formed 
compact object, either by stellar material originally close to the surface 
of the imploding core and left behind by the explosion-generating outgoing 
shock, or by more distant stellar matter falling back after its passage 
(De R\'ujula~1987). As observed in microquasars, each time part of the 
disk falls abruptly onto the compact object, a pair of CBs made of {\it 
ordinary plasma} are emitted with high bulk-motion Lorentz factors, 
$\gamma$, in opposite directions along the rotation axis, wherefrom matter 
has already fallen onto the compact object, due to lack of rotational 
support. The $\gamma$-rays of a single pulse in a GRB are produced as a CB 
coasts through the SN glory --the SN light scattered by the SN and pre-SN 
ejecta. The electrons enclosed in the CB Compton up-scatter (Shaviv \& Dar 
1995) glory's photons to GRB energies.

Each pulse of a GRB corresponds to one CB. The emission times of the 
individual CBs reflect the chaotic accretion process and are not 
predictable. At the moment, neither are the characteristic baryon number 
and Lorentz factor of CBs, which can be inferred from the analysis of GRB 
afterglows (DDD02; DDD03; DD04). Given this information, two other 
`priors' (the typical early luminosity of a core-collapse supernova and 
the typical density distribution of the parent star's wind-fed circumburst 
material), and a single extra hypothesis (that the wind's column density 
in the `polar' directions is significantly smaller than average) all 
observed properties of the GRB pulses can be successfully predicted 
without the introduction of any ad-hoc parameters (Dar \& De 
R\'ujula~2004, thereafter DD04).

The spectral energy density, $F_\nu(t)$, of the X-ray emission of a GRB 
has two phases. The first is very rapidly declining X-ray emission and 
dominated by the late-time tail of the GRB pulses (DD04) and/or by
line emission and thermal bremsstrahlung from the CBs (DDD02;
Dado et al.~2006). In a second 
phase, synchrotron radiation from swept-in ISM electrons spiraling in the 
CBs' enclosed magnetic field takes over. This second phase has a 
`plateau': a shallow time-dependence lasting until the CBs decelerate 
significantly in their collisions with the interstellar medium, after 
which $F_\nu(t)$ bends into an asymptotic power-law decline $\sim 
t^{-1.6}$. On this basis we were able to predict (DDD02) the `canonical' 
behaviour of X-ray AGs, observed by Swift (Dado et al.~2006).

In the CB model, three times characterize the evolution of $F_\nu(t)$.
The first is the time at which synchrotron radiation begins to dominate.
The second is the time at which the {\it injection bend} of the electron
energy spectrum within a CB crosses a particular frequency,
and corresponds to a strongly chromatic change in $F_\nu(t)$
(DDD02; DDD03). The third time, the typical deceleration  
time, corresponds to an achromatic change in $F_\nu(t)$ (DDD02)
and is the subject of this paper.

\section{The deceleration time}
\label{dectime}

Let $\theta\!=\!{\cal{O}}$(1 mrad) be the typical viewing angle of an 
observer of a CB that moves with a typical Lorentz factor 
$\gamma\!=\!{\cal{O}}(10^3)$.  Let $\delta\!=\!{\cal{O}}(10^3)$ be the 
corresponding Doppler factor:
\begin{equation}
\delta \equiv {1\over\gamma\,(1-\beta\, \cos\theta)}
                       \simeq  {2\, \gamma
                       \over 1+\gamma^2\, \theta^2}\; ,
\label{delta}
\end{equation} 
where the approximation is excellent 
for  $\theta\ll 1$ and  $\gamma \gg 1$. 

A CB is assumed
to contain a tangled magnetic field in equipartition with the ISM protons
that enter it. As it ploughs through the ionized ISM, a CB gathers and
scatters its constituent protons. The re-emitted protons exert an inward
pressure on the CB, countering its expansion. 
Let $n\simeq n_p\simeq n_e$ be the number density in a dominantly 
hydrogenic  ISM. In the approximation of
isotropic re-emission in the CB's rest frame and a constant $n$,
one finds that within minutes of
observer's time $t$, a CB reaches a nearly-constant `coasting'
radius $R$. Subsequently, $\gamma(t)$ obeys:
\begin{equation}
[({\gamma_0/ \gamma})^{3+\kappa}-1]+
(3-\kappa)\,\theta^2\,\gamma_0^2\,
 [(\gamma_0/\gamma)^{1+\kappa} - 1]
= t/t_0;\,\,\,\,\,t_0 \equiv {(1+z)\, N_b\over
(6+2\kappa)\,c\, n\,\pi\, R^2\, \gamma_0^3}\; , 
\label{cubic}
\end{equation}
with $\kappa=1(0)$ depending on whether the re-emitted ISM particles are a 
small  (large) fraction of the intercepted ones.
This dichotomy is too small to detect in the study of AGs, but the case
$\kappa=1$, which we adopt, is favoured by the CB model of Cosmic 
Rays (Dar \& De R\'ujula, 2006).

To specify the CB-model's prediction for an end-time,
$T'_a$, paraphrasing  the one defined by Willinger et al.~(2007) and
Nava et al.~(2007), let us derive the time at which the X-ray AG is smaller by
a factor of two than the extrapolation from its previous shallow behaviour,
and let us refer to the typical parameters of observed GRBs, for which
$\gamma_0\,\theta\!\sim\! 1$, and $\delta(t)\!\approx\!\gamma(t)$
in the shallow phase (DDD02). The  X-ray AG, as we shall recall in Section \ref{XAGenergy}, Eq.~(\ref{Fnuobser}),
behaves as $F_{_{\rm X}}(t)\!\propto\!\gamma(t)^{6.4}$, so that we are
demanding that $[\gamma(t)/\gamma_0]^{6.4}\!\simeq\!1/2$. Insert this into the
left hand side of Eq.~(\ref{cubic}), with $\kappa\!=\!1$ and 
$\theta\gamma_0\!=\!1$, to conclude that
the typical end-time is $t\!=\!1.026\,t_0$. Nava et al.~(2007) correct this time for the
cosmological redshift; according to Eq.~(\ref{cubic}), its predicted value is:
\begin{eqnarray}
T'_a &=&
{T_a\over 1+z}\approx 1.026\,{t_0\over 1+z}\sim (1.4\times 10^3\, {\rm s})\,V_T\nonumber\\
V_T&=&\left[{\gamma_0\over 10^3}\right]^{-3}\,
\left[{n\over 10^{-2}\, {\rm cm}^{-3}}\right]^{-1}\,
\left[{R\over 10^{14}\,{\rm cm}}\right]^{-2}\,
\left[{N_b\over 10^{50}}\right],
\label{Ta}
\end{eqnarray}
where we have normalized to typical CB-model parameters and the
`variability' around them is governed by the combination of
parameters $V_T$ (DDD03; Dar \& De R\'ujula 2006).

\section{The isotropic energy in the prompt GRB}
\label{promptE}

In the CB model  the isotropic  (or spherical equivalent)
energy, $E_\gamma^{\rm iso}$, of a GRB, is (DD04): 
\begin{eqnarray} 
E_\gamma^{\rm iso} &\simeq& 
{\delta^3\, L_{_{\rm SN}}\,N_{_{\rm CB}}\,\beta_s\over 6\, c}\,
                      \sqrt{\sigma_{_{\rm T}}\, N_b\over 4\, \pi}\sim
                      (2.8\! \times\! 10^{53}\;{\rm erg})\,V_E,
                     \nonumber\\
                      V_E&\equiv&{\delta^3\over 10^9}\,
{L_{_{\rm SN}}\over L_{_{\rm SN}}^{\rm bw}}\,{N_{_{\rm CB}}\over 4.5}\,
\beta_s\sqrt{ N_b\over 10^{50}}\; ,
\label{eiso} 
\end{eqnarray} 
where $L_{_{\rm SN}}$ is the mean SN optical luminosity just 
prior to the ejection of  CBs, $N_{_{\rm CB}}$ is the number of CBs in 
the jet, $N_b$ is their mean baryon number, $\beta_s$ is the comoving early
expansion velocity of a CB (in units of $c/\sqrt{3}$), 
and $\sigma_{_{\rm T}}$ is the Thomson cross section. The early SN luminosity required to 
produce the mean isotropic energy, $E_\gamma^{\rm iso}\!\sim\! 4\!\times\! 10^{53}$ 
erg, of ordinary long GRBs is 
$L_{_{\rm SN}}^{\rm bw}\!\simeq\! 5\!\times\! 10^{42}\, {\rm erg\, 
s^{-1}}$, the estimated early luminosity of SN1998bw. All quantities in
Eq.~(\ref{eiso}) are normalized to their typical CB-model values.
For $N_{_{\rm CB}}$ we took the result of a recent careful analysis
of the number of significant peaks in a GRB light curve (Schaefer~2006) 
rather than the one we previously adopted ($N_{_{\rm CB}}=6$, DD04).

Nava et al.~(2007) choose to present their results in terms of the prompt
isotropic energy in the 15-150 keV domain. To restrict the `bolometric'
result of Eq.~(\ref{eiso}) to a fixed-energy bracket, we must recall the
prediction of our model for a GRB's spectral shape (DD04). 
The photons of the glory's light that a GRB Compton-upscatters
have a thin-bremsstrahlung spectrum 
$dN/dE_i\propto(1/E_i)^\alpha\,{\rm Exp}[-E_i/T_i]$, with 
$\alpha\sim 1$ and $T_i\sim 1$ eV. The bulk of these electrons are
comoving with the CB and Lorentz- and Doppler- boost the
target light to a spectrum of the same shape, and 
 `temperature':
\begin{equation}
(1+z)\,T\sim {4\over 3} \;T_i\;{\gamma\;\delta}\,\langle 1+\cos\theta_i\rangle
\sim 1.3\,{\rm MeV}\,\langle 1+\cos\theta_i\rangle,
\label{Teff}
\end{equation}
where $\theta_i$ is the angle of incidence of a glory's photon
into the CB, in the SN rest system.
A very tiny fraction of the moving electrons is due to `knock-on',
or is `Fermi-accelerated' within the CB, in both cases to a
spectrum (in the CB's rest frame) $dN_e/dE_e\propto E^{-p}$,
with $p\simeq 2.2$. The complete prompt $\gamma$ spectral
distribution, upscattered by the CB's comoving and knock-on
electrons (DD04),  is:
\begin{equation}
E\,{dN_\gamma\over dE}
\propto
\left({T\over E}\right)^{\alpha-1}\; e^{-E/T}+b\;
(1-e^{-E/T})\;
{\left(T\over E\right)}^{p/ 2}
\label{totdist}
\end{equation} 
For $b={\cal{O}}(1)$, and $\alpha$ and $p$ in their 
expected range, the above spectrum is uncannily similar to the
phenomenological `Band' spectrum (Band et al.~1993). The `peak energy'
of the prompt spectrum is:
\begin{equation}
 (1+z)\,E_p\simeq {\gamma\,\delta\, \epsilon_p\over 2}\simeq
 (500\;{\rm keV})\; {\gamma\,\delta\over 10^6}\,
{\epsilon_p\over 1\;\rm eV}\; .
 \label{eobs}
\end{equation}

The average redshift of Swift GRBs, and of the ones discussed here,
is $\langle z \rangle=2.8$.
The fraction $f_\gamma$ of the bolometric $E_\gamma^{\rm iso}$ lying in the
15-150 keV range is the ratio of $\int\! E\,dN_\gamma$ in the range 
$\langle 1+z \rangle\times$(15-150) keV, to
the same integral from 0 to $\infty$: $f_\gamma\simeq 0.106$ for
all parameters at their central values and $\langle 1+\cos\theta_i\rangle=1/2$
(a semitransparent glory, DD04). Our  prediction is then:
\begin{equation}
E_\gamma^{\rm iso}[15\!-\!150 \,{\rm keV}]
=f_\gamma\,E_\gamma^{\rm iso}
=(2.9\! \times\! 10^{52}\;{\rm erg})\,V_E,
\label{eisointerval}
\end{equation}
with $V_E$ as in Eq.~(\ref{eiso}). As the rest of our results, $f_\gamma$
is computed for `typical parameters', corresponding to a relatively
large $E_p$ value and the concomitant large bolometric corrections.
For many post-Swift GRBs the bolometric correction would be smaller.

\section{The isotropic energy in the X-ray plateau phase}
\label{XAGenergy}

In the plateau phase and thereafter, the CB-model's AG is due to synchrotron
emission by the electrons continuously entering a CB from the interstellar medium
(ISM) it sweeps. Above observer's radio frequencies, and in the CB's rest system,
the synchrotron radiation has a (normalized)  spectral shape (DDD03):
\begin{eqnarray} 
 \nu\,{dn_\gamma\over d\,\nu} &\propto&  
f_{\rm sync}(\nu,t) = 
{K(p)\over \nu_b(t)}{[\nu/\nu_b(t)]^{-1/2}\over
\sqrt{1+[\nu/\nu_b(t)]^{(p-1)}}}
\nonumber \\  
  K(p)&\equiv&
{\sqrt{\pi}\over  2\, \Gamma\left[{2\, p-1\over 2(p-1)}\right]
                                   \, \Gamma\left[{ p-2\over 2(p-1)}\right]} 
                                   \approx {p-2\over 2\,(p-1)},
\label{sync} 
\end{eqnarray}
where the  {\it `injection bend'} frequency $\nu_b$ corresponds to the energy,
$E_b=m_e\,c^2\,\gamma(t)$, at which ISM electrons enter the CB at the time when
its Lorentz factor is $\gamma(t)$. The predicted
bend frequency $\nu_b$ ($\nu_b^{\rm obs}$) in the CB's (observer's) frame is:
\begin{equation}
{\delta\over 1+z}\,\nu_b = \nu_b^{\rm obs}
\simeq {(5.9\times 10^{15}\;{\rm Hz}) \over 1+z}\, 
{[\gamma(t)]^3\, \delta(t)\over 10^{12}}\,
\left[{n\over 10^{-2}\;{\rm cm}^3}\right]^{1/2}\,.
\label{nubend}
\end{equation}
The typical frequency in the parenthesis is equivalent to an energy of 3.9 eV.
This is always below the X-ray domain, so that the corresponding X-ray spectrum
has a $\sim\!\nu^{-1.1}$ shape. But,
occasionally, at times of order 1 day or less, the observed optical frequencies
are above $\nu_b^{\rm obs}$, so that the optical spectrum varies from
$\sim\!\nu^{-0.5}$ to $\sim\!\nu^{-1.1}$, producing a chromatic break occurring in the
optical AG but not in the X-ray one (DDD03).

In a CB's rest frame, the energy flux density in the optical to X-ray domain 
is:
\begin{equation} 
F_{_{\rm CB}}[\nu,t] \simeq
               \eta\, \pi\,R^2\, n\, m_e\, c^3\, \gamma(t)^2  
 \,f_{\rm sync}(\nu,t), 
\label{Fnucb} 
\end{equation} 
where $\eta$ is the fraction of ISM electrons that enter the CB
and radiate there the bulk of their incident energy, and $f_{\rm sync}$ is
as in Eq.~(\ref{sync}).
The AG spectral energy density $ F_{\rm obs}$
seen by a cosmological observer at a redshift $z$, is:
\begin{equation} 
F_{\rm obs}[\nu,t]\simeq N_{_{\rm CB}}\;{
 (1+z)\,\delta(t)^3 
                    \over 4\, \pi\, D_L^2}\, 
F_{_{\rm CB}}\left[{(1+z)\,\nu\over\delta(t)},{\delta(t)\,t\over 1+z} 
\right]\propto \gamma(t)^{2.3}\,\delta(t)^{4.1}
\, , 
\label{Fnuobser} 
\end{equation} 
where 
$F_{_{\rm CB}}$
is as in Eq.~(\ref{Fnucb}), and $D_L$ is the luminosity distance.
As announced in the derivation of Eq.~(\ref{Ta}), 
$F_{\rm obs}\propto\gamma^{6.4}$
in the shallow phase of the X-ray afterglow.

With use of the spectrum of Eqs.~(\ref{sync},\ref{nubend}) we can define a fraction $f$
 of the spectral energy in the 15-150 keV X-ray band. For $p=2.2$:
\begin{equation}
f\equiv K[2.2] \int_{15\,{\rm keV}}^{150\,{\rm keV}}d\nu\;{[\nu_b^{\rm obs}]^{0.1}\over \nu^{1.1}}\approx 0.14\,\left[{3.8\over \langle1+z\rangle}\;
{\gamma^3\,\delta\over 10^{12}}\;
\left({n\over 10^{-2}\,{\rm cm^{-3}}}\right)^{1/2}\right]^{0.1}\; .
\label{f}
\end{equation}
Gathering the above results, we obtain for the equivalent isotropic energy
per unit observer time in the specified X-ray range:
\begin{eqnarray}
&&F_{_{\rm X}}[15\!-\!150 \,{\rm keV}]=f\,F_{_{\rm X}}=
N_{_{\rm CB}}\,\eta\,f\,\pi\,R^2\,n\,m_e\,c^3\,\gamma^2\delta^4/(1+z)
\simeq (9.7 \times 10^{47}\,{\rm erg\,s^{-1}})\; \eta\,V_F
\nonumber\\
&&V_F(t)=\left[{\gamma(t)\over 10^3}\right]^{2.3}\,
\left[{\delta(t)\over 10^3}\right]^{4.1}\,
\left[{3.8\over \langle 1+z \rangle}\right]^{1.1}\,
\left[{n\over 10^{-2}\, {\rm cm}^{-3}}\right]^{1.05}\,
\left[{R\over 10^{14}\,{\rm cm}}\right]^{2}\,
\left[{N_{_{\rm CB}}\over 4.5}\right]\, ,
\label{FX}
\end{eqnarray}
where we used $\langle 1+z \rangle=3.8$ for the average redshift of
GRBs detected by Swift.

The corresponding integrated X-ray energy in the plateau is:
\begin{equation}
I_{_{\rm X}}^{\rm iso}\simeq {1\over 2}\,F_{_{\rm X}}(t\,=\,0)\,T_a=
(6.8\times 10^{50}\,{\rm erg})\,V_T\,V_F(0),
\label{IX}
\end{equation}
where we took $\eta\simeq 1$, $V_T$ and $V_F$ are as in Eqs.~(\ref{Ta})
and (\ref{FX}), and
the factor 1/2 reflects the fact that (as can be seen in a plot
which is not a logarithmic) most of the plateau, as defined here,
extends in the domain where the AG light curve has a value
$\sim 1/2$ of its initial value.

\section{Results}
\label{Results}

\subsection{Afterglow versus prompt bolometric energies}
\label{GRBvsAG}

A simple and crucial test of models of GRBs is the predicted
ratio of the bolometric energy in a GRB's afterglow
up to the end of the plateau phase (essentially all of the
AG's energy) and the total energy in the GRB's prompt
$\gamma$ rays. According to Eqs.~(\ref{FX}) and (\ref{IX}), 
the CB-model expectation is:
\begin{eqnarray}
R[{\rm AG/GRB}] &=& 
{3\;\delta_0\over 4\;\gamma_0}\,\sqrt{\pi\,N_b\over\sigma_{_{\rm T}}}\;
m_e\,c^3\,{\eta\over L_{_{\rm SN}}\,\beta_s}=0.08\;V_R\, ,
\nonumber\\
V_R &=&{2\over 1+\theta^2\,\gamma_0^2}\;{\eta\over\beta_s}\,
\sqrt{N_b\over 10^{50}}\;
{L_{_{\rm SN}}^{\rm bw}\over L_{_{\rm SN}} }\; .
\label{ratio}
\end{eqnarray}
This ratio is rather `clean': it establishes a link between 
the late and prompt emissions which is
 independent of the number of CBs, of their radii, of the density
 of the ISM in which they travel,
and weakly dependent on their baryon number. It very naturally
explains why the observed ratios are typically
of the order of a few percent.

\subsection{Central and typical values}
\label{CandT}

Our main results are Eq.~(\ref{Ta}) for the time ending the
shallow X-ray AG decay,  Eq.~(\ref{eisointerval}) for 
$E_\gamma^{\rm iso}[15\!-\!150 \,{\rm keV}]$,
a GRB's prompt isotropic-equivalent energy in the specified interval, and
Eq.~(\ref{IX}) for the isotropic energy, $I_{_{\rm X}}^{\rm iso}$,
 in that energy interval 
of the X-ray AG,
integrated in time up to the end of the plateau. The `typical' parameters
underlying these results are based on the analysis of pre-Swift GRBs,
and reflect the domain wherein GRBs, in the past and with less performing
satellites, it was easiest to detect GRBs.

\begin{figure}[]
\hspace{-0cm}
\epsfig{file=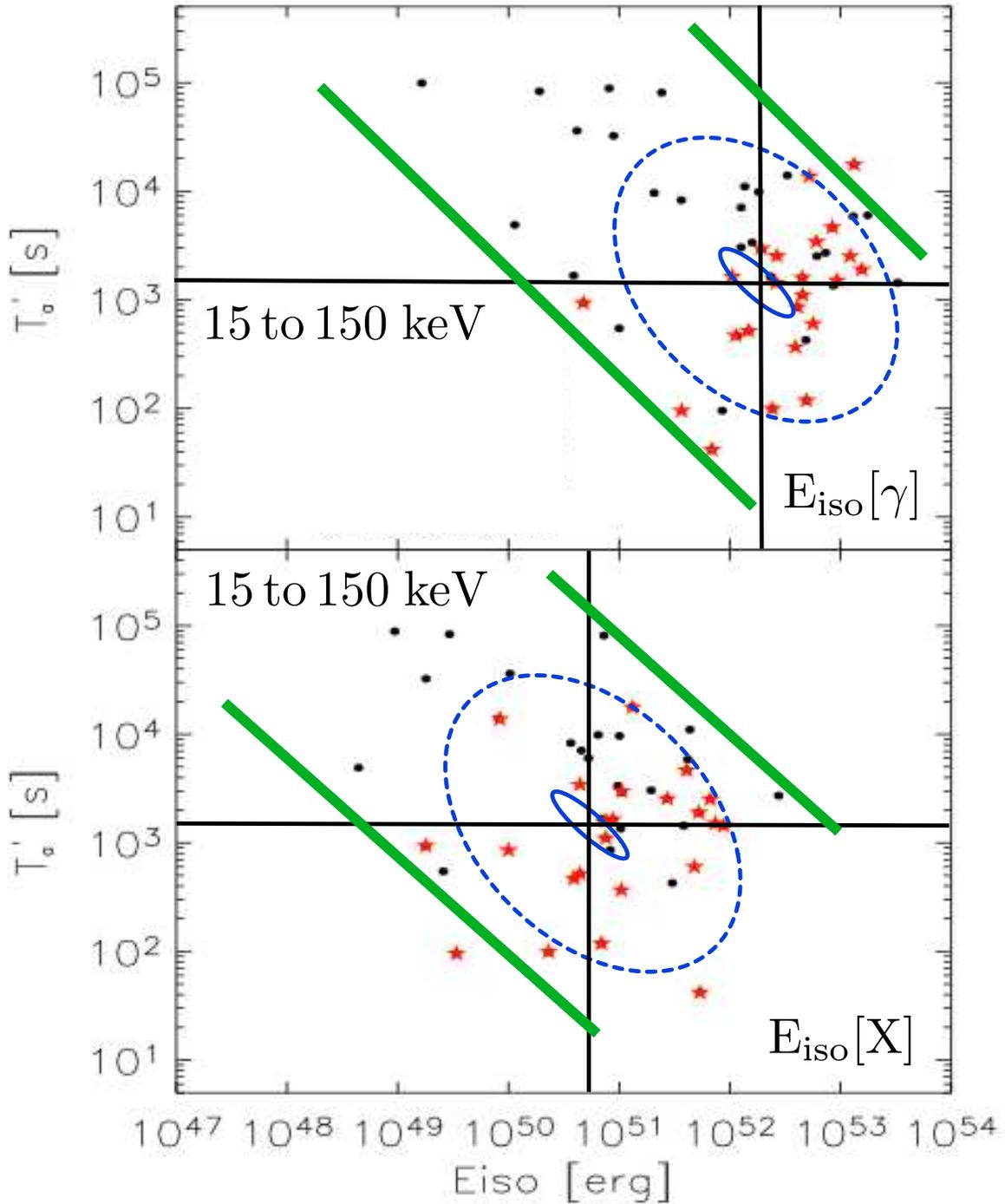,width=19cm}
\vspace{-8cm}
\caption{The central and typical values, discussed in Section 
\ref{CandT},
for the time ending the X-ray plateau, plotted against the isotropic energies
of the prompt $\gamma$ rays and of the X-ray AG up to that time.
The data are those gathered by Nava et al.~(2007). 
The crossing lines are the 
predictions of Eqs.~(\ref{Ta},\ref{eisointerval},\ref{IX}). 
Within the small ellipse, the 
parameters $\gamma_0$ and $\delta_0$ are allowed to range in the
small domain in which most pre-Swift GRBs gathered. The larger
ellipse allows for the relevant combinations of the other case-by-case
parameters to vary by about one order of magnitude. The thick
(green) lines
are the expected trend of the correlations.}
\label{f2} 
\end{figure}

The predicted central expectation for $T'_a$ is shown in both
parts of Fig.~\ref{f2} as a horizontal line. The central expectations for
$E_\gamma^{\rm iso}[15\!-\!150 \,{\rm keV}]$ and $I_{_{\rm X}}^{\rm iso}$
are the vertical lines in the upper and lower part of the figure, respectively.
The `sweet spot' drawn as the
small ellipses corresponds to letting $\gamma$ and $\delta$ vary in
the observed narrow domain wherein most pre-Swift GRBs lied
(DDD02; DD04). The larger
dotted ellipses are drawn by allowing the combinations of variability
parameters, $V$ in Eqs.~(\ref{Ta}, \ref{eisointerval}, \ref{IX}),
vary by about an order of magnitude around the small
ellipse. As expected, most Swift GRBs with relatively large and measured
peak energy (the stars) and relatively large prompt and AG isotropic
energies, are within the dotted ellipses. 
Most of the extra points (the dots)
may in the past have been classified as XRFs (they have low `peak energy',
$E_p$). We discuss them in the next subsection.

The green lines in Fig.~\ref{f2} show the correlations expected
for typical parameters. For them, the $\gamma$ and $\delta$ dependences
of the relevant quantities are $T'_a\propto \gamma_0^{-3}$, 
$E_\gamma^{\rm iso}\propto \delta_0^3$ and 
$I_X^{\rm iso}\propto \gamma_0^{-0.7}\,\delta_0^{4.1}$, so that
$T'_a\propto 1/E_\gamma^{\rm iso}$ and 
$T'_a\propto (I_X^{\rm iso})^{-3/3.4}$ for $\delta_0\approx\gamma_0$.

\subsection{Distributions and correlations}
\begin{figure}[]
\centering
\epsfig{file=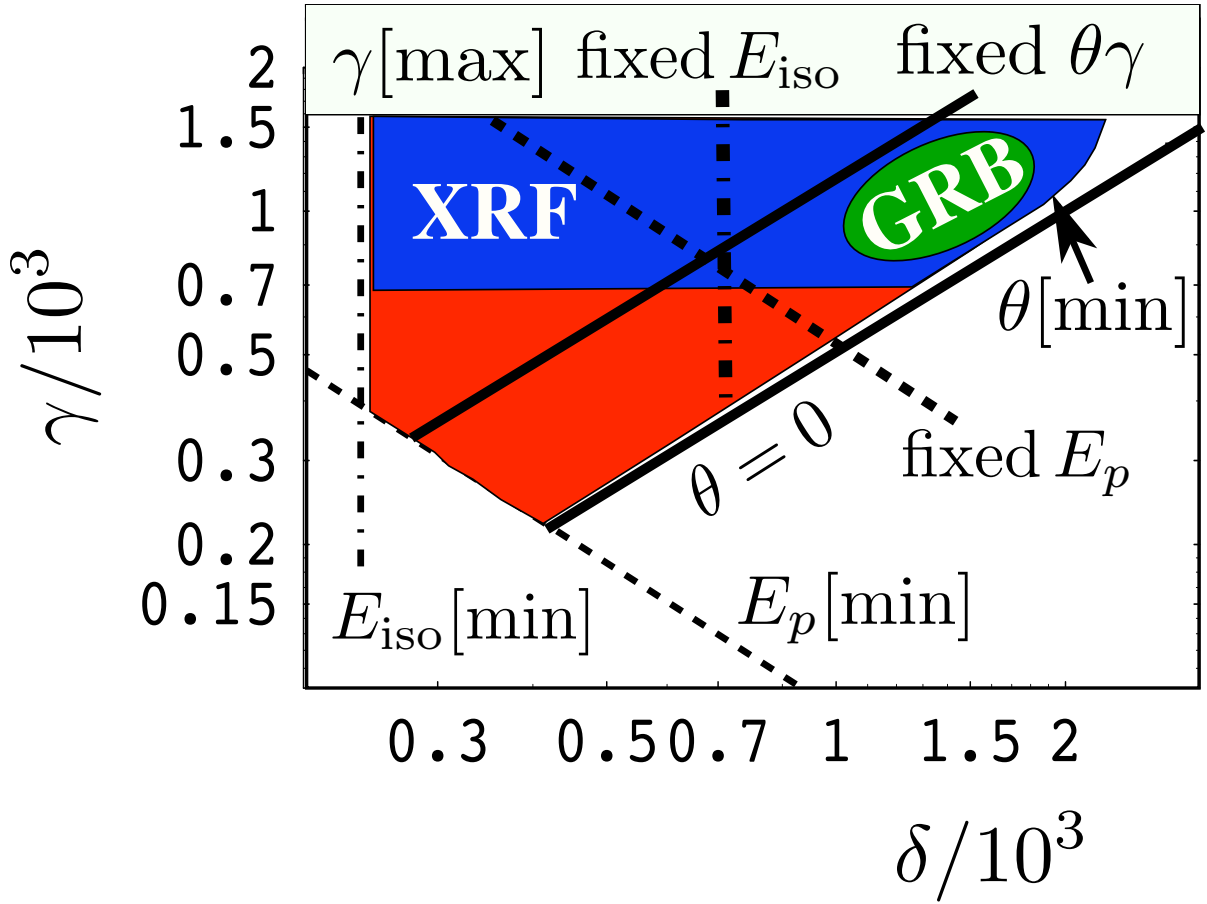,width=16cm}
\caption{The domain of $(\delta,\,\gamma)$ values. The green elliptical
spot labeled `GRB' is the area wherein most pre-Swift GRBs were observed.
The region to its left has relatively small (large) $\delta$
($\theta$) values, corresponds to relatively small $E_\gamma^{\rm iso}$
and $E_p$, and is labelled `XRF'.
The blue horizontal band is limited above and below, reflecting the
narrow distribution of the $\gamma$ values of observable CBs
(DDD02; DD04). }
\label{f3} 
\end{figure}

In the CB model, XRFs are the same as GRBs, but observed at
a relatively large angle, $\theta$ (or a particularly small $\gamma$),
implying a small $\gamma\,\delta$ (DD04; Dado et al.~2004). Thus, XRFs 
have a relatively small 
spectral peak energy  and a small prompt isotropic energy. 
The explicit proportionality factors in the relations
$E_p\propto\gamma\,\delta$
and $E_\gamma^{\rm iso}\propto\delta^3$ are given by
Eqs.~(\ref{eobs},\ref{eiso}). Consider them fixed
at their typical values.
The typical $(\gamma,\,\delta)$ domain of observable GRBs is 
then the one shown in 
Fig.~\ref{f3}. The observed values of $\gamma$ are fairly
narrowly distributed around $\gamma\!\sim\!10^3$ (DDD02, DD04),
as in the blue strip of the figure. The $(\gamma,\,\delta)$ domain
is also limited by a minimum
observable isotropic energy or fluence (both $\propto\,\delta^3$), by
a minimum observable peak energy, and by the line
$\theta=0$ or  by a line corresponding to
a minimum fixed $\theta$, if one takes into account that phase space for
observability diminishes as $\theta\to 0$. 
The elliptical `sweet spot' in  Fig.~\ref{f3}
is the region wherein GRBs are most easily detectable, particularly
in pre-Swift times.  X-ray Flashes populate the region labeled
XRF in the figure, above the fixed $\gamma\theta$ line or to the
left of the fixed $E_p$ line. We  interpret most of the dotted points in
Figs.~\ref{f3} and \ref{f4} as cases for which $\gamma$ and $\delta$
lie in the `XRF domain' of Fig.~\ref{f3}.

The continuous red lines in Fig.~\ref{f4} are the contours of the blue 
domain of Fig.~\ref{f3}, projected into the $[E_\gamma^{\rm iso},\,T'_a]$ 
plane (top) and the $[E_X^{\rm iso},\,T'_a]$ plane (bottom). The 
projectors' are the corresponding functions of $\gamma$ and $\delta$, 
e.g.~$T'_a(\gamma,\delta)$, as in Eq.~(\ref{Ta}). The dotted red lines are 
drawn by `moving' the red contour about its `central' position in the 
planes, by approximately one order of magnitude, once again to reflect the 
variability of parameters other than $\gamma$ and $\theta$ (or $\gamma$ 
and $\delta$). The dotted red lines satisfactorily describe the location 
and distribution of the Swift data. The green lines are the predicted 
trend of the correlations, which for the ensemble of the data (stars and 
points) interpolates between two power laws, as in Fig.~\ref{f1}d, and as 
discussed in detail for this and many other correlations in Dado et 
al.~2007.

 \section{Conclusions}
 
We have analised data on two afterglow observables (the time ending the 
shallow decay of X-ray AGs and the integrated isotropic energy up to that 
point) as well as a prompt-GRB observable (the isotropic energy). To do 
so, we have simply reported the theoretical expectations of the CB model, 
developed in previous papers. The predictions include the magnitudes of 
these quantities, the explicit dependence on the parameters that govern 
their case-by-case variability, and the spectral shapes of the prompt and 
afterglow phases.
 
\begin{figure}[]
\centering
\vbox{
\hbox{
 \epsfig{file=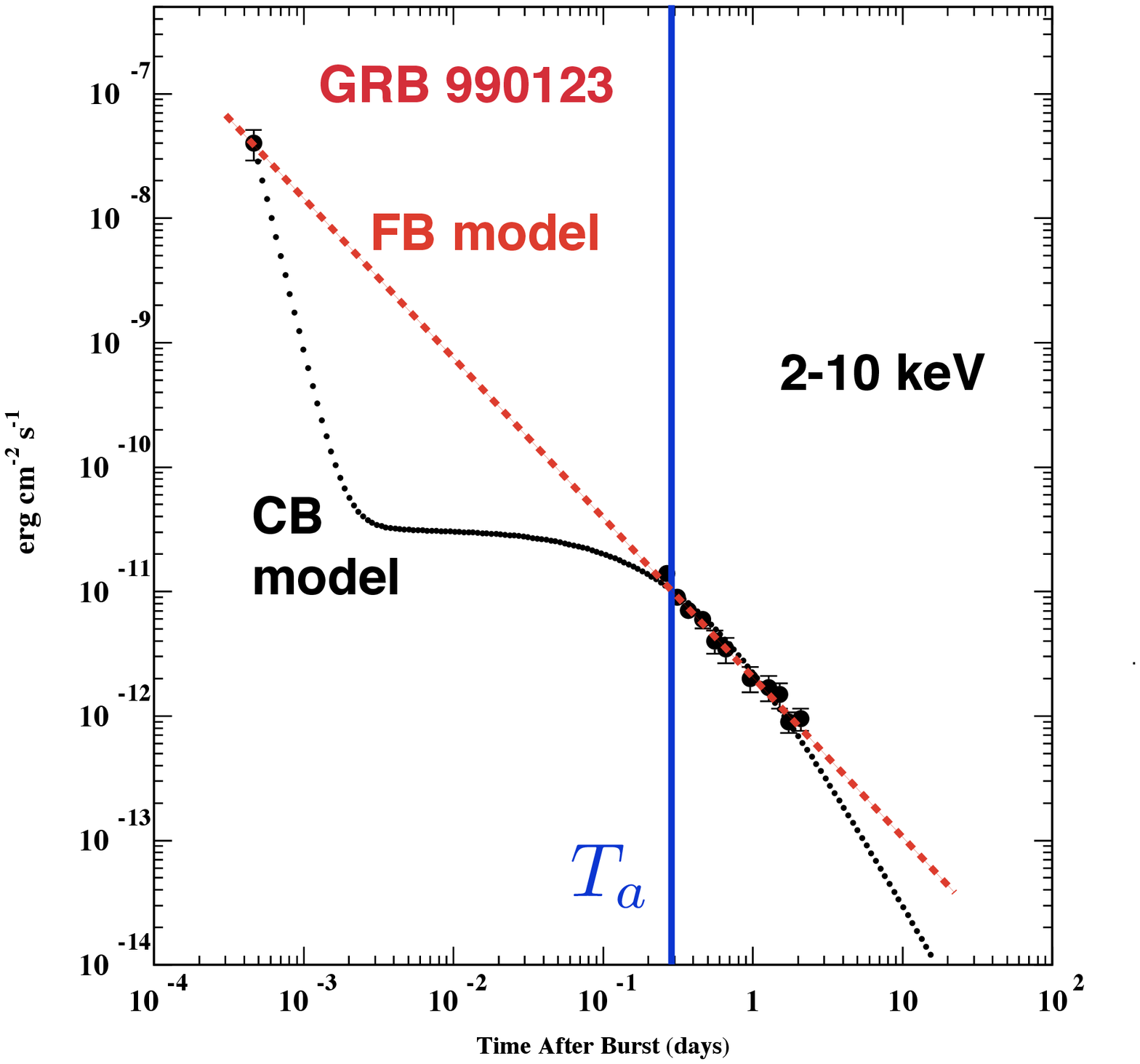,width=8cm}
 \epsfig{file=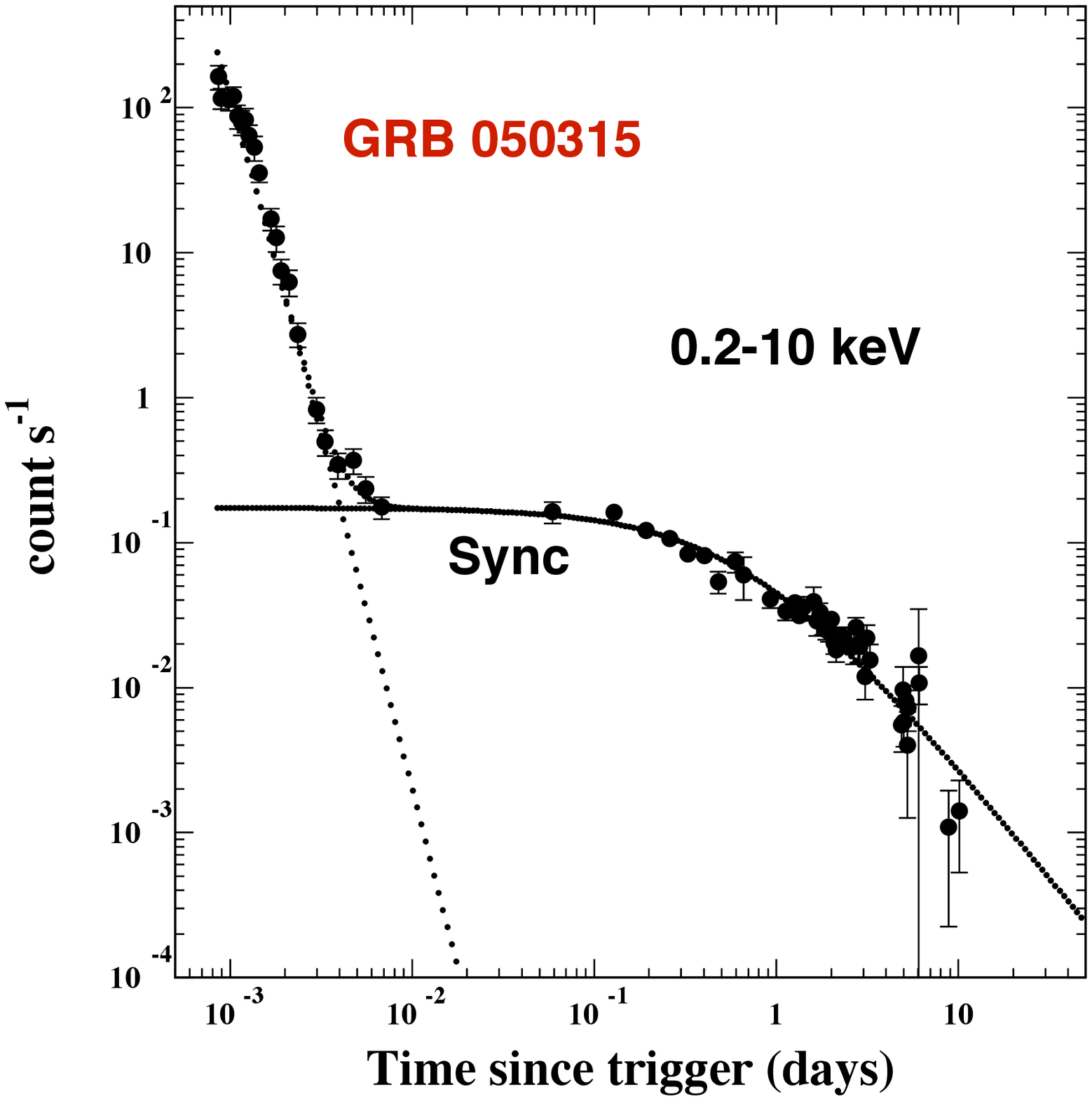,width=8cm }
}}
\centering
\vbox{
\hbox{
\epsfig{file=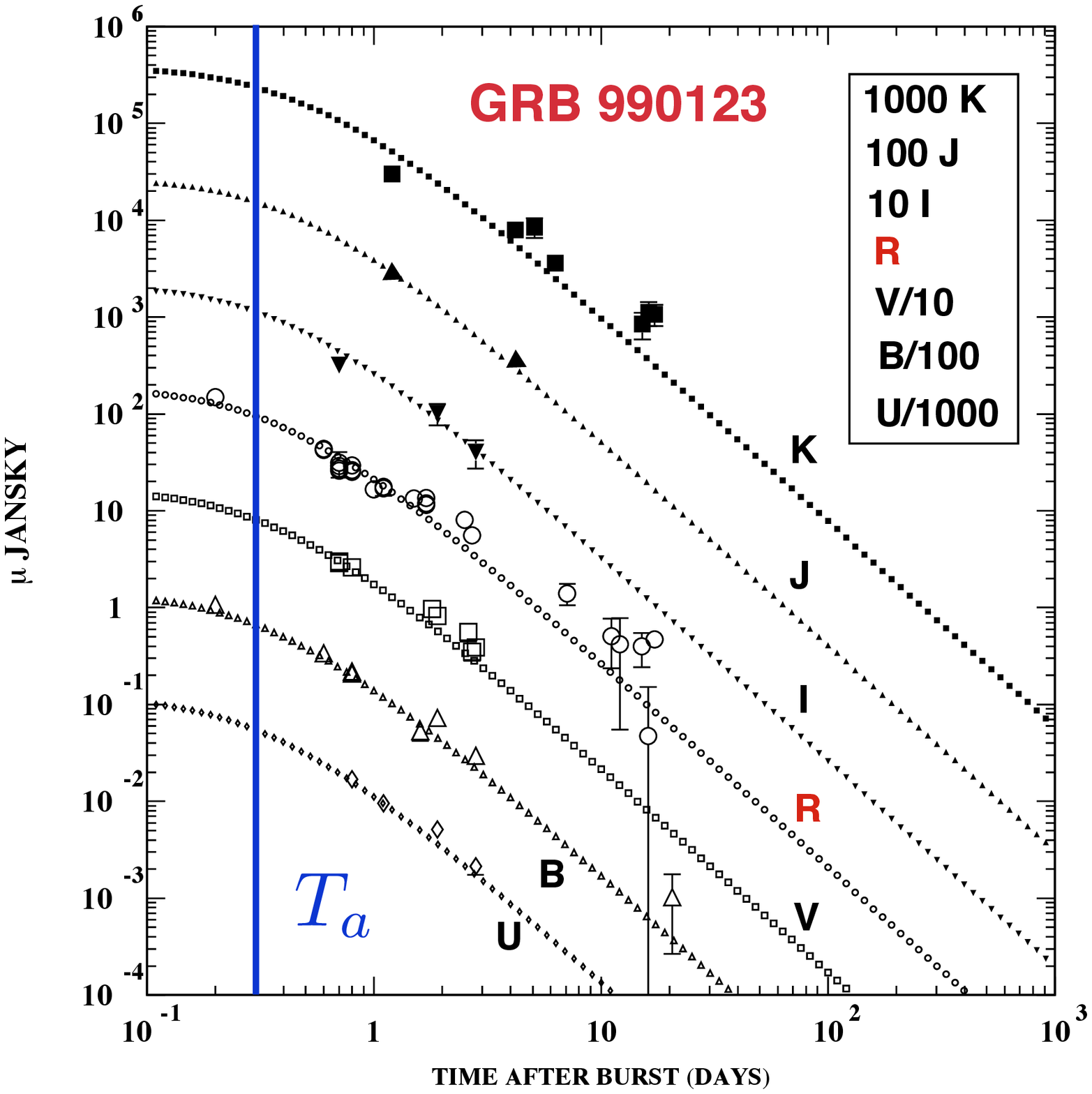,width=8.cm}
\epsfig{file=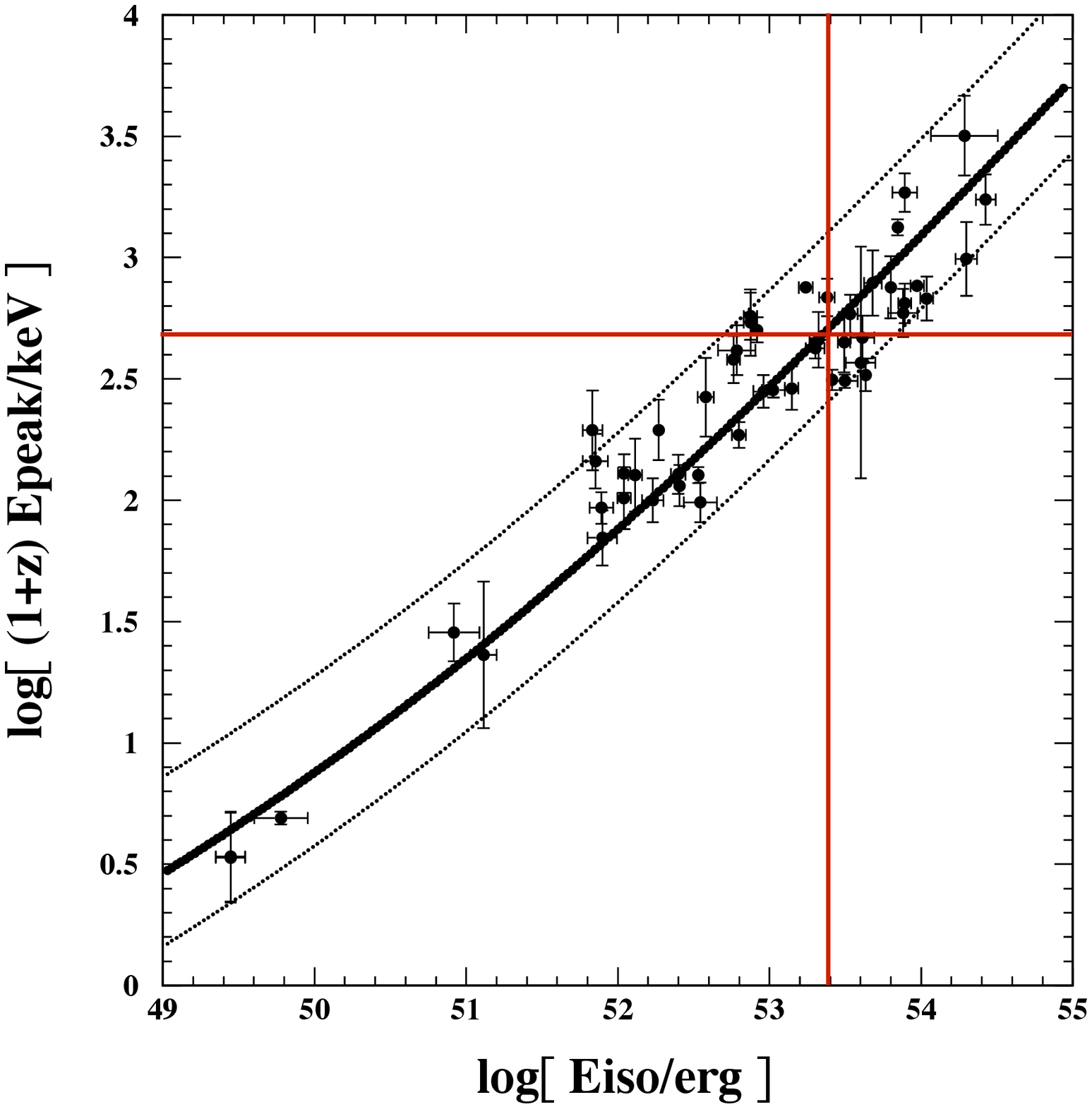,width=8.cm}
}}
\vspace*{8pt}
\caption{{\bf Top left}: Pre-Swift predictions for the 2-10 keV X-ray AG
in the CB (DDD02) and fireball (Maiorano et al.~2005) models, compared to
data for GRB 990123. $T_a$ is the time ending the plateau phase.
{\bf Top right}: Broad band optical data on GRB 990123, fit in the CB model
(DDD03). The evolution is achromatic all the way up to the X-ray energies.
{\bf Bottom left}: Comparison between the CB model prediction and the 
canonical 0.2-10 keV X-ray light curve of GRB  050315 (Vaughan et al.~2006)
{\bf Bottom right}: The $(E_p,\,E_\gamma^{\rm iso})$ correlation, compared
with its predicted trend in the CB model (Dado et al.~2007). The crossed
red lines are the  predicted typical or average values, see
Eqs.~(\ref{eiso},\ref{eobs}).}
\label{f1}
\end{figure}

\begin{figure}[]
\hspace{-0cm}
\epsfig{file=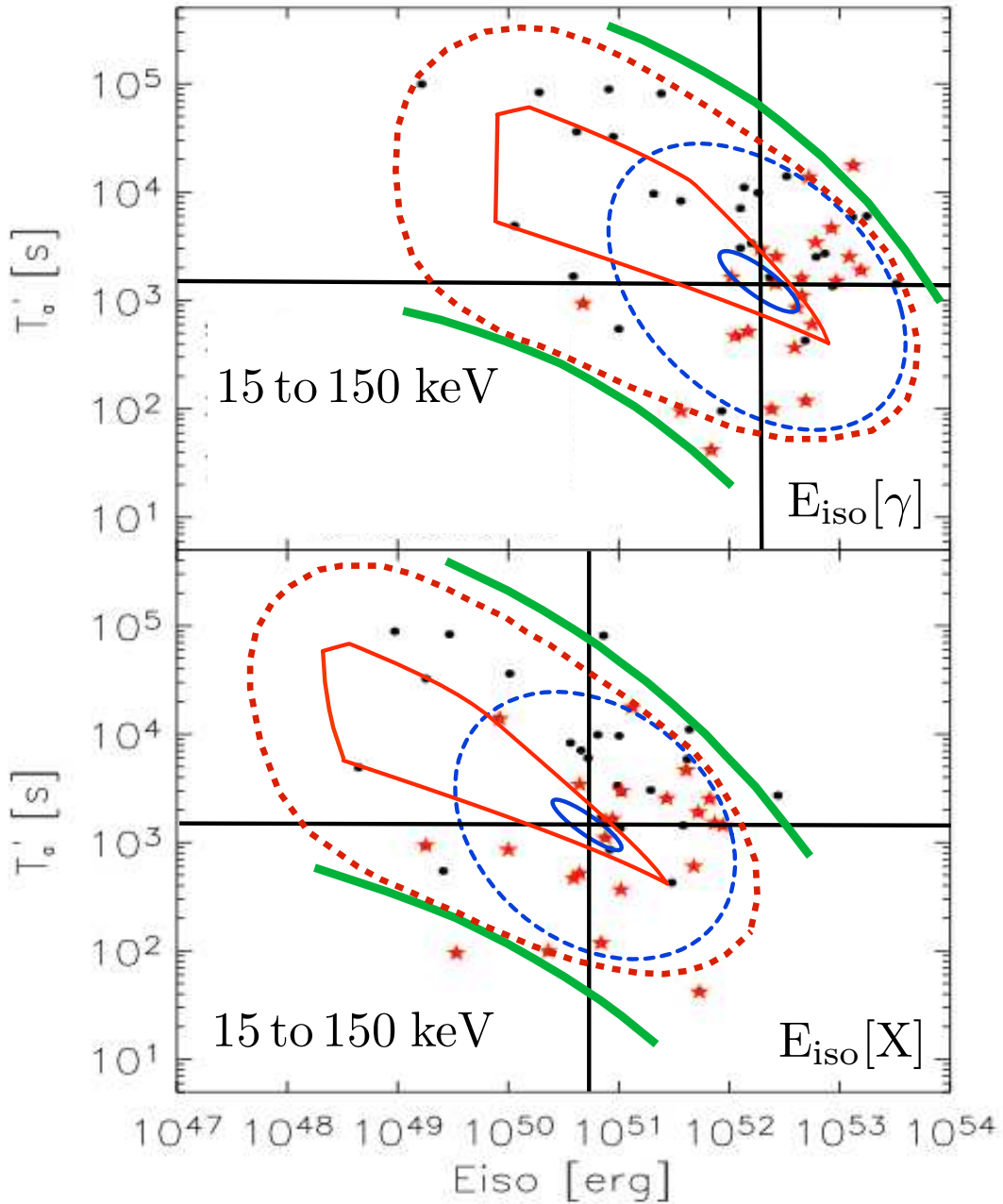,width=15cm}
\caption{The time ending the X-ray plateau, plotted against the 
isotropic energies
of the prompt $\gamma$ rays and of the X-ray AG up to that time. The data,
crossing lines and ellipses are as in Fig.~\ref{f2}. The red line in
the top figure is the projection of the contour of the blue 
$(\gamma,\,\delta)$ domain of Fig.~\ref{f3} onto the 
$(T'_a,\,E_\gamma^{\rm iso})$ plane. It encompasses the area
in which GRBs and XRFs are expected to lie, for all parameters set
to their central values, but for $\gamma$ and $\delta$. The red dashed 
contour is obtained by letting the rest of the relevant combinations of
parameters vary by about one order of magnitude. The green thick lines
show the expected trend of the correlations. The lower figure is
built in the same way, in the $(T'_a,\,E_X^{\rm iso})$ plane. }
\label{f4} 
\end{figure}

The results can best be summarized by looking at Fig.~\ref{f4}, whose data 
points are those in the corresponding figure in Nava et al.~(2007) who 
discuss, among others, the same subject. The predictions of the CB model, 
predating the Swift data, are in excellent agreement with the 
observations. The data are centered and distributed as expected. Their 
correlations are also the expected ones, though, since the data points do 
not span a large number of orders of magnitude, they are not as remarkable 
as for other correlations, such as the one in Fig.~\ref{f1}a.
 
In our opinion, the main novelty in the paper of Nava et al.~(2007) is the 
discussion of correlations between prompt-GRB and GRB-afterglow 
observables. These test the ensemble and coherence of a GRB-model's 
ingredients. In the CB model the prompt $\gamma$ rays are of Compton 
origin, while the AG light is dominated by synchrotron radiation. Unlike 
fireball models based on very different physics, the CB model never had an 
`energy crisis' (see e.g.~Piran 2000) in the relation 
between the total energies in the prompt and afterglow phases, or a 
problem with the prompt spectrum (see, e.g.~Ghisellini 2003). The time ending 
the shallow afterglow phase is a `deceleration time' of the cannon-balls, 
unrelated to the opening jet-angle of fireball models. The CB model 
provides very simple, predictive and successful descriptions of the prompt 
(DD04, Dado et al. 2007) and afterglow phases (DDD02; DD03; 
Dado et al.~2006; Dado et al.,~in preparation). We are not surprised that 
the model is also successful in the detailed description of the 
distributions of prompt and afterglow observables, and of their 
correlations.

{\bf Acknowledgment:} This 
research was supported in part by the Asher Space Research Fund  
at the Technion.

 
\end{document}